\documentclass[11pt]{article}
\usepackage{xspace,amssymb,amsmath,helvet,graphicx}
\begin{document}
\newcommand{\dee}{\,\mbox{d}}
\newcommand{\naive}{na\"{\i}ve }
\newcommand{\eg}{e.g.\xspace}
\newcommand{\ie}{i.e.\xspace}
\newcommand{\pdf}{pdf.\xspace}
\newcommand{\etc}{etc.\@\xspace}
\newcommand{\PhD}{Ph.D.\xspace}
\newcommand{\MSc}{M.Sc.\xspace}
\newcommand{\BA}{B.A.\xspace}
\newcommand{\MA}{M.A.\xspace}
\newcommand{\role}{r\^{o}le}
\newcommand{\signoff}{\hspace*{\fill} Rose Baker \today}
\newenvironment{entry}[1]%
{\begin{list}{}{\renewcommand{\makelabel}[1]{\textsf{##1:}\hfil}%
\settowidth{\labelwidth}{\textsf{#1:}}%
\setlength{\leftmargin}{\labelwidth}
\addtolength{\leftmargin}{\labelsep}
\setlength{\itemindent}{0pt}
}}%
{\end{list}}
\title{Application of some new heavy-tailed survival distributions\\Rose D. Baker\footnote{Centre for Operational Research and Applied Statistics,
University of Salford, UK.} Email r.d.baker@salford.ac.uk}  
\maketitle
\begin{abstract}
Some new survival distributions are introduced based
on a generalised exponential function. This class of distributions includes 
heavy-tailed generalisations of exponential, Weibull and gamma distributions.
Properties of the distributions are described, and R code is available for
computation of pdf, quantiles, inverse quantiles, random numbers, etc.
A use of these distributions for robust inference is suggested, and this is exemplified
with a Monte-Carlo study.
\end{abstract}
\section*{Keywords}
Exponential distribution, Lomax distribution, Weibull distribution,  arcsinh transformation, heavy tail.
\section{Introduction}
This paper introduces some new survival distributions related to the arcsinh transformation.
The arcsinh transformation of the random 
variable was introduced by Johnson in 1949 (see \eg  Johnson {\em et al}, 
1994, Chapter 1) and more recently developed by Jones and Pewsey (2009). It is possible to create new distributions
by transforming the random variable $X$ to $\nu\sinh^{-1}(X/\nu)$, thus creating a distribution with an extra parameter $\nu$ that reduces to the original distribution as $\nu \rightarrow \infty$.
Here however, the arcsinh transformation is used in a looser way, by defining a generalised logarithm 
\[\ln_{\nu}(x)=\nu\sinh(\ln(x)/\nu)=\nu\{x^{1/\nu}-x^{-1/\nu}\}/2\]
and exponential function 
\[\exp_{\nu}(x)=\exp(\nu\sinh^{-1}(x/\nu))=(\sqrt{1+(x/\nu)^2}+x/\nu)^{-\nu}.\]
Fresh distributions are then created by replacing exponentials and logarithms that occur in familiar distributions with their generalisations. This method leads to heavy-tailed versions of these
familiar distributions; this paper focuses on survival distributions.

Why should a statistician be interested in these new distributions, given that so many survival distributions have already been invented?
The characteristic feature of the new distributions is that they are very similar to their `parent' distribution in the body, and only show real divergence
in the tail. To peek ahead, figures \ref{figlb} and \ref{figla} show a generalised exponential distribution, along with the original distribution and the Lomax distribution,
a `type 2' Pareto distribution that is basically the Pareto distribution modified to start at $x=0$. It can be seen indeed that the generalised distribution
is more similar to the exponential in the body of the curve than the Lomax distribution, whilst being heavy-tailed.

This behaviour is sometimes useful. For example, we may believe that a distribution is roughly exponential, but want the analysis to be able to cope with outliers.
Using a generalised distribution with a high value of $\nu$ will give very similar results to fitting an exponential when there are no outliers, but
will cope with them if they are present. 

When setting up Bayesian prior distributions, heavy tails are needed if we wish to obviate the effects of prior-data clashes (\eg, O'Hagan and Pericchi, 2011).
Expert opinion may be that a distribution is exponential; however the wily Bayesian statistician will add a heavy tail that will safeguard against clashes
but will not greatly distort the expert opinion that has been elicited. Also, when using any form of Monte-Carlo method for Bayesian computations such as MCMC,
a heavy-tailed proposal distribution is {\em de rigeur}. In importance sampling for example, if the sampling distribution is too short-tailed, there will be very few random variables
generated in the tail, but they will have huge weights. This makes the method very inefficient, and so heavy-tailed sampling distributions are needed.
Distributions that resemble the `parent' distribution in the body of the distribution are more efficient.

In the following sections, heavy-tailed generalisations of the exponential, Weibull and gamma distributions are introduced and their properties
derived. A Monte-Carlo study of the use of these distributions in robust inference is described. First, there is a brief note on mathematical methods.
\section{A note on Mathematical methods}
To evaluate integrals throughout, for determining unknown constants and evaluating moments,
it is helpful to change variables to $q(x)=\{C+S\}^{-2}$, where $C=\sqrt{1+(x/\nu)^2}$, $S=x/\nu$. Then $0 < q < 1$ and $\dee q/\dee x=-(2/\nu)q/C$.
Since $C-S=1/(C+S)$, we have that $S=x/\nu=(q^{-1/2}-q^{1/2})/2$, $C=(q^{-1/2}+q^{1/2})/2$. Hence a general integral such as
\begin{multline}\int_0^y f_x(x)x^{\gamma}C(x)^{\delta}\dee x=\\
\int_{q(y)}^1 f_q(q) (\nu/2)^{\gamma+1}(1/2)^{\delta+1}q^{-(\gamma+\delta+1)/2}(1-q)^{\gamma}(1+q)^{\delta+1}\dee q,\label{eq:maths}\end{multline}
where $f_x(x)=f_q(q)$. Since in practice $\delta$ is -1 or 0 and $f_q(q)$ is a power of $q$ multiplied by a power of $x$, all required integrals can be done as sums of incomplete beta functions.

Throughout, $f(x)$ denotes a pdf, $S(x)$ the survival function, and $F(x)$ the distribution function, so that $S(x)=1-F(x)$. The corresponding hazard function
$h(x)=-\dee \ln S(x)/\dee x$. We write $f_a(x)$, $S_a(x)$, $F_a(x)$ and $h_a(x)$ for the corresponding arcsinh-related distributions.

The `standard' distributions are presented, and these can be trivially 
generalised by adding two parameters so that $X \rightarrow (X-\eta)/\tau$, when the support of the distribution becomes $X \ge \eta$. Then $S(x) \rightarrow S((x-\eta)/\tau)$
and $f(x) \rightarrow (1/\tau)f((x-\eta)/\tau)$.

\section{An exponential-like distribution}
The usual long-tailed version of the exponential distribution is the Pareto; here the Lomax distribution is
discussed instead, as this is a version of the Pareto defined for $X > 0$. The survival function $S(x)=(1+x/\nu)^{-\nu}$ and the pdf is
$f(x)=(1+x/\nu)^{-\nu-1}$.

Clearly at least two generalised exponential distributions could be defined, 
by taking the pdf $f_a(x) \propto \exp_\nu(-x)$, or the survival function 
$S_a(x)=\exp_\nu(-x)$. The latter is simpler, and is explored further 
here; the former choice is given briefly in a later section.
 
Now $S_a(x)=\{\sqrt{1+(x/\nu)^2}-x/\nu\}^\nu=\{\sqrt{1+(x/\nu)^2}+x/\nu\}^{-\nu}$. 
It can also be written as $S_a(x)=\exp(-\nu\sinh^{-1}(x/\nu))$. The 
pdf is 
\begin{equation}f_a(x)=S_a(x)/\sqrt{1+(x/\nu)^2},\label{eq:type1}\end{equation} so the hazard $h_a(x)=1/\sqrt{1+(x/\nu)^2}$.
The squaring of $x$ in the hazard function shows how the distribution can resemble the exponential closely except in the tail.
The median $m$ is $m=\nu\sinh\{\ln(2)/\nu\}$, and in general the quantile 
function $Q$ such that $F_a(Q(p))=p$ is given by $Q=\nu\sinh(-\ln(1-p)/\nu)$. 
This result can be used for random number generation; if $U$ is a uniformly-distributed 
random number, then $\nu\sinh(-\ln(U)/\nu)$ is a random variable from 
this distribution. The mode is zero as for the exponential, so the pdf monotonically decreases with $x$.
 
The moments are readily obtained on applying (\ref{eq:maths}).
\[\text{E}(X)=\nu^2/(\nu^2-1)\] for $\nu > 1$,
\[\text{E}(X^2)=2\nu^2/(\nu^2-4)\] for $\nu > 2$,
so that the variance
\[\sigma^2=\frac{\nu^2(\nu^4+2)}{(\nu^2-1)^2(\nu^2-4)}\] for
$\nu > 2$. The third moment is
\[\text{E}(X^3)=\frac{6\nu^4}{(\nu^2-1)(\nu^2-9)}\]
for $\nu > 3$
and the skewness is then
\[\gamma=\frac{2\nu(\nu^6+2\nu^4+6\nu^2+15)(\nu^2-4)^{1/2}}{(\nu^2-9)(\nu^4+2)^{3/2}}.\]
In the next section, a 
general formula for moments, which need not be integer, is given.

For completeness, the moment generating function defined for $t < 0$ is
\[M(t)=\nu\int_0^\infty \exp\{\nu(t\sinh(\theta)-\theta)\}\dee
\theta=\frac{\nu\pi}{\sin(\nu\pi)}[{\bf J}_\nu(-\nu
t)-J_\nu(-\nu t)],\] where ${\bf J}$ is the Anger function and $J$ the 
Bessel function (Gradshteyn and Ryzhik 3.482 (3)).

Using the same transformation of $x/\nu=\sinh(\theta)$, the entropy is
\[\nu\int_0^\infty(\ln \cosh \theta-\nu\theta)\exp(-\nu\theta)\dee \theta,\]
and on integrating the first term by parts and using Gradshteyn and Ryzhik 3.541 (7),
the entropy is $1-1/\nu+(1/2)\psi(\frac{\nu+2}{4})-(1/2)\psi(\frac{\nu}{4})$,
where $\psi$ denotes the Euler psi (digamma) function.
 
\begin{figure}
\centering
\makebox{\includegraphics{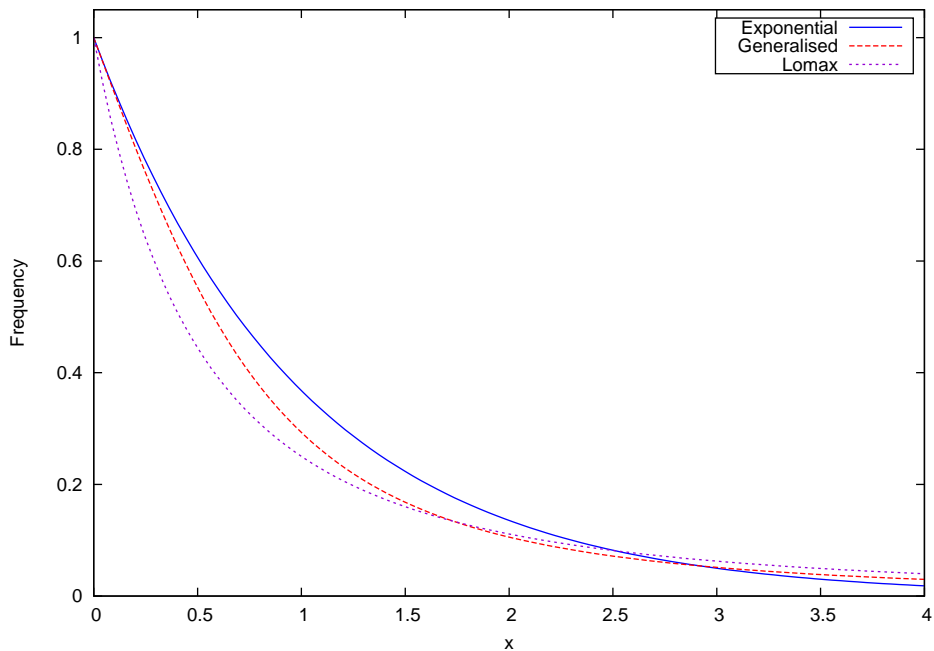}}
\caption{\label{figlb} The exponential pdf,  the modified pdf from equation (\ref{eq:type1}),
and the Lomax pdf. Here $\nu=1$.}
\end{figure}

\begin{figure}
\centering
\makebox{\includegraphics{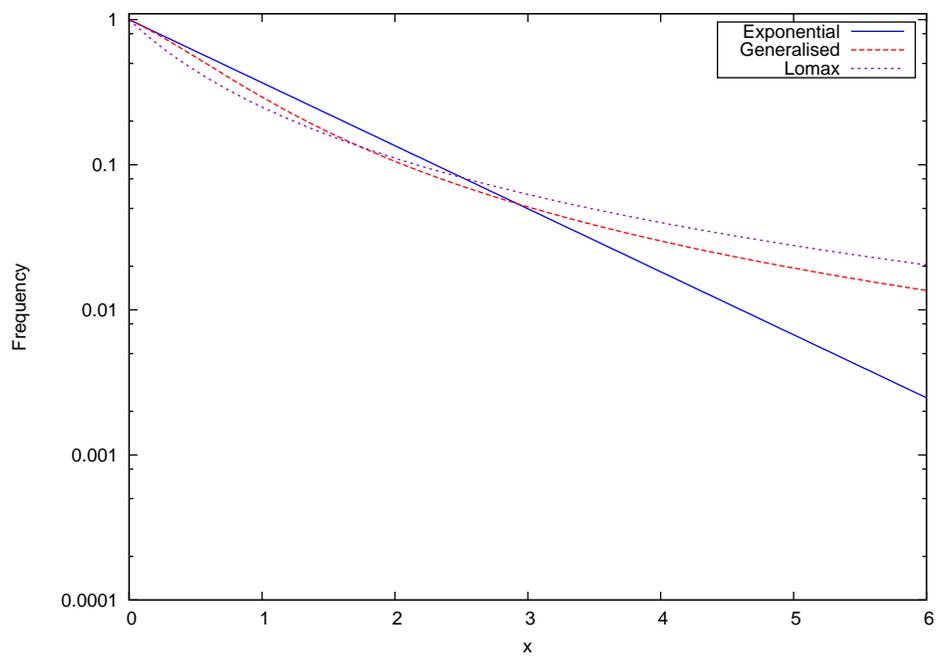}}
\caption{\label{figla} The exponential pdf,  the modified pdf from equation (\ref{eq:type1}),
and the Lomax pdf plotted on a logarithmic scale. Here $\nu=1$.}
\end{figure}

This distribution differs from the Lomax in that it is more exponential 
in the body of the distribution, but with the same power-law tail behaviour. 
Writing the Lomax survival function as $(1+x/\nu)^{-\nu}$, on expanding 
the logarithm, one sees that $\ln S(x) \simeq -x+(1/2)(x^2/\nu+\cdots$, 
whereas for the proposed distribution, $\ln S_a(x) \simeq -x-(\nu/6)(x/\nu)^3+(3\nu/40)(x/\nu)^5\cdots$. 
Hence the deviation from exponentiality is $O(x^3)$, compared with $O(x^2)$
for the Lomax distribution. Figure \ref{figlb} shows how similar the generalised distribution is in the body of the curve, 
and the dissimilarity of the Lomax pdf, while figure \ref{figla} shows the power-law tail. Here $\nu=1$.

\section{A heavy-tailed Weibull distribution}
The Weibull distribution is usually made long-tailed by giving the scale factor an inverse gamma distribution, when we obtain
the Singh-Maddala (Burr type 12) distribution where $S(x)=(1+x^\beta/\nu)^{-\nu}$.

To derive a corresponding arcsinh-related distribution we write the survival function
\[S_a(x)=\exp_\nu(-x^\beta)=\{\sqrt{1+(x^\beta/\nu)^2}+x^\beta/\nu\}^{-\nu},\]
so generalising the generalised exponential to a generalised Weibull. Here the scale factor is 
set to unity. This can also be written $S_a(x)=\exp(-\nu\sinh^{-1}(x^\beta/\nu))$. The pdf is
\[f_a(x)=\beta x^{\beta-1}\frac{\{\sqrt{1+(x^\beta/\nu)^2}+x^\beta/\nu\}^{-\nu}}{\sqrt{1+(x^\beta/\nu)^2}},\]
For $\beta > 1$ and $\nu > 0$ this gives a mode with a long tail, the 
hazard function being
\[h_a(x)=\frac{\beta x^{\beta-1}}{\sqrt{1+(x^\beta/\nu)^2}}.\]
This initially increases for $\beta > 1$ and then decreases to zero, 
where $h_a(x) \simeq \beta\nu/x$ in the tail. The hazard function is a maximum at $x=\nu^{1/\beta}(\beta-1)^{1/2\beta}$ for $\beta > 1$.
 
The moments are tractable. Changing variable to $y=x^\beta$ and applying (\ref{eq:maths}) 
we readily obtain
\[\text{E}(X^n)=(\nu/2)^{1+n/\beta}\text{B}((\nu-n/\beta)/2,1+n/\beta),\]
where $B$ denotes the Beta function. Clearly the $n$th moment only exists 
if $n < \beta\nu$. Fractional and inverse moments where $n$ is not integer 
can also be found. The formula for the median is
\[m=\{\nu\sinh(\ln (2)/\nu)\}^{1/\beta}.\]
The formula for the mode is derived by differentiating the logged pdf,
\[\ln f_a(x)=(\beta-1)\ln(x)-\nu\ln(C+S)-\ln(C)+\text{const}\]
to obtain
\[\dee \ln f_a(x)/\dee x=0=\frac{\beta-1}{x}-\frac{\beta x^{\beta-1}}{C}-\frac{\beta x^{2\beta-1}/\nu^2}{C^2}.\]
Multiplying out the denominator, 
\[\beta x^\beta C=(\beta-1)C^2-\beta x^{2\beta}/\nu^2,\]
and on squaring, a quadratic is obtained for $(x^\beta/\nu)^2$.
The mode for $\beta > 1$ is then
\[x_m=\{\frac{2(\beta-1)^2\nu^2}{(\nu\beta)^2+2(\beta-1)+\sqrt{(\nu\beta)^4+4(\nu\beta)^2\beta(\beta-1)}}\}^{1/2\beta}.\]
This reduces to $\{\frac{(\beta-1)}{\beta}\}^{1/\beta}$ as $\nu\rightarrow\infty$ as it must.
 
Random numbers can be generated by generating numbers $Y$ from the generalised exponential
distribution and then setting $X=Y^{1/\beta}$. 
\section{A long-tailed gamma distribution}
The usual distribution that would be used here is the F distribution, a type of Pearson type 6 (\eg Johnson, Kotz and Balakrishnan, p 382), also known as the compound gamma distribution.
This has pdf 
\[f(x)=\frac{\nu^{-1}(x/\nu)^{\beta-1}}{\text{B}(\nu,\beta)(1+x/\nu)^{\nu+\beta}}.\]
The corresponding arcsinh-related distribution is obtained by replacing the exponential as was done for the exponential distribution, when we obtain 
from the gamma pdf $f(x)=x^{\beta-1}\exp(-x)/\Gamma(\beta)$ the pdf
\[f_a(x) \propto \frac{x^{\beta-1}\{\sqrt{1+(x/\nu)^2}+x/\nu\}^{-\beta-\nu+1}}{\sqrt{1+(x/\nu)^2}},\]
where on integrating using (\ref{eq:maths}), finally
\[f_a(x)=(2/\nu)^\beta\frac{x^{\beta-1}\{\sqrt{1+(x/\nu)^2}+x/\nu\}^{-\beta-\nu+1}}{\sqrt{1+(x/\nu)^2}\text{B}(\frac{\nu}{2},\beta)}.\]
The survival function is the incomplete beta function
\[S_a(x)=\text{B}(\{\sqrt{1+(x/\nu)^2}+x/\nu\}^{-2};\frac{\nu}{2},\beta).\]
The moments for $ n < \nu$ are
\[\text{E}(X^n)=(\nu/2)^n \frac{\text{B}(\frac{\nu-n}{2},\beta+n)}{\text{B}(\frac{\nu}{2},\beta)}.\]
The mode for $\beta > 1$ is
\[x=\frac{2^{1/2}\nu(\beta-1)}{\{B+\sqrt{B^2+4(\nu+2\beta-3)(\nu+1)(\beta-1)^2}\}^{1/2}},\]
where $B=\nu^2+(\beta-1)(2\nu-\beta+3)$. The derivation as for the long-tailed Weibull distribution is by differentiating the logged pdf and setting the derivative to zero,
when on multiplying out the denominator and squaring, a quadratic equation for $(x/\nu)^2$ is obtained. Random numbers are best obtained by generating random numbers
from the corresponding F distribution, and using the rejection method.
Thus considering the ratio $f_a(x)/f(x)$, $C(x)+S(x) > 1+x/\nu$, but $C(x) < 1+x/\nu$. Writing $g(x)=\frac{1+x/\nu}{C(x)}$, the maximum of $g$, $g_{\text{max}}$ is slightly less than 1.42.
The rejection method is guaranteed to work if we generate random numbers from the F distribution, and accept them with probability
\[p(x)=\frac{(1+x/\nu)^{\nu+\beta}}{\sqrt{1+(x/\nu)^2}(C+S)^{\nu+\beta-1}g_\text{max}}.\]

\section{Another exponential-type distribution}
Here instead of taking the survival function to be $\exp_{\nu}(-x)$ we take the pdf $f_a(x) \propto \exp_{\nu}(-x)$.
In fact it is easier to rescale $x$ and to take $f_a(x)=kq^{\nu+1}$, where $q=\sqrt{1+(x/\nu)^2}-x/\nu$, so that the distribution is defined if $\nu > 0$.

To find the constant of proportionality $k$, we apply (\ref{eq:maths}) to the pdf
so that $k=(\nu+2)/(\nu+1)$ and the pdf is
\begin{equation}f_a(x)=(\nu+2)/(\nu+1)(\sqrt(1+(x/\nu)^2+x/\nu)^{-\nu-1}.\label{eq:type2}\end{equation}

The properties follow similarly, with the survival function 
\[S_a(x)=\frac{1}{2(\nu+1)}\{\nu q(x)^{\nu+2}+(\nu+2)q(x)^\nu\}.\]
The moments are: $\text{E}(X)=\frac{\nu^2(\nu+2)}{(\nu-1)(\nu+1)(\nu+3)}$ for $\nu > 1$,
$\text{E}(X^2)=\frac{2\nu^2}{(\nu-2)(\nu+4)}$ for $\nu > 2$ and so on. This distribution is messier than the exponential-like distribution derived
by taking the survival function as a generalised exponential function, and is not considered further.

There are of course many other survival distributions that could be `arcsinhed', but the most important ones have been given.
\section{Probabilistic derivation}
The Lomax distribution can be generated by making the scale factor $w=1/\tau$ of the exponential distribution
a gamma-distributed random variate. We naturally ask what pdf $g(w)$ would convert the exponential distribution
into the generalised exponential distribution, so that 
\begin{equation}\int_0^\infty \exp(-x w)g(w)\dee w=k(x/\nu+\sqrt{1+(x/\nu)^2})^{-\nu}.\label{eq:lap}\end{equation}
The generalised exponential is the Laplace transform of $g(w)$, from which using the table of Laplace transforms in Schiff
(1999), p 214, we have $g(w) \propto J_{\nu}(w)/w$, where $J$ is the Bessel function of the first kind.
Bessel functions oscillate in sign, so $g(w)$ is not a pdf, and hence \ref{eq:lap} does not admit a probabilistic interpretation.
The generalised exponential distribution cannot be derived from the exponential distribution by making the scale a random variable, the way one converts an exponential distribution to the Lomax,
the Weibull to the Burr type 12 (Singh-Maddala) distribution, or the gamma to the F distribution. 

\section{Examples}
The use of these new distributions in inference is illustrated.
In practice, inference would typically be made for regression coefficients, so $\tau$ might be regressed on some covariates. It is true that there is always a
modelling choice as to what is the best parameter to regress on covariates. For example, for the Weibull distribution should one regress the mean (which is a 
complicated function of parameters) on covariates or the mode? Here, we simply seek to estimate $\tau$ for an exponential distribution in the presence of outliers.

To illustrate the use of the new distributions in robust analysis, samples of exponentially-distributed random variables were generated,
and one or two outliers added. The data were fitted to the Lomax and generalised exponential distributions, to estimate  the scale parameter $\tau$.
Table ref{tab1} shows the estimated values of $\tau$ and $\nu$, and the logged likelihood.
It can be seen that the generalised exponential usually gives the better fit to the data. Table \ref{tab2} shows the error in the estimate of $\tau$, relative
to the sample mean omitting the outliers. It can be seen that the generalised exponential fit gives a smaller error than the \naive fit of
an exponential distribution, and also smaller than the fit obtained using the Lomax distribution. The use of the Lomax distribution is not always preferable to
simply ignoring the possibility of outliers, but fitting the generalised exponential considerably reduces the error.

Of course, instead of fitting a fat-tailed distribution that can downweight outliers, one could attempt to detect outliers and
omit them from the dataset; however, this procedure is hazardous and robustifying the analysis by fitting a heavy-tailed distribution that automatically downweights outliers without discarding them
is preferable.
\begin{table}[h]
\begin{tabular}{|l|l|c|c|c|c|c|c|} \hline
&&\multicolumn{3}{c|}{Lomax}&\multicolumn{3}{c|}{Generalised exponential}\\ \hline
Sample & Outliers&$-\ell$&$\hat{\tau}$&$\hat{\nu}$&$-\ell$&$\hat{\tau}$&$\hat{\nu}$\\ \hline
10&0&-&.999&-&&&\\ \hline
10&1&16.19&.379&.693&16.69&.516&600\\ \hline
10&2&21.33&.455&.638&21.92&.596&.542\\ \hline
100&0&-&.8865&-&&&\\ \hline
100&1&100.482&.797&4.51&98.85&.866&2.56\\ \hline
100&2&107.36&.800&3.63&105.69&.879&2.16\\ \hline
1000&0&-&.9848&-&&&\\ \hline
1000&1&1001.6&.9468&18.1&999.1&.9713&5.86\\ \hline
1000&2&1010.2&.9427&14.89&1007.4&.972&5.22 \\ \hline
\end{tabular}
\caption{\label{tab1}Results of fitting a sample of exponentially-distributed random numbers with outliers added with the Lomax and generalised exponential distributions.
Minus log-likelihood, estimated $\tau$ and degrees of freedom are shown. The first outlier is 20, the second is 10.}
\end{table}
\begin{table}[h]

\begin{tabular}{|l|l|c|c|c|c|c|c|} \hline
Sample & Outliers&Error: ignore&Error: Lomax&Error: generalised exponential\\ \hline
10&1&1.7274&.62&.483\\ \hline
10&&2.3335&.544&.403 \\ \hline
100&1&.1892&.0895&.0205 \\ \hline
100&2&.2767&.0865&.0075 \\ \hline
1000&1&.019&.038&.0135 \\ \hline
1000&2&.028&.0421&.0128 \\ \hline
\end{tabular}
\caption{\label{tab2}Results of fitting a sample of exponentially-distributed random numbers with outliers added with the Lomax and generalised exponential distributions.
Errors from ignoring the outlier, fitting a Lomax distribution and fitting the generalised exponential are shown. The first outlier is 20, the second is 10.}
\end{table}
\section{Conclusions}
Some new survival distributions have been introduced, which are heavy-tailed generalisations of the exponential, Weibull and gamma distributions.
They are analogous to the Pareto (strictly, type 2 Pareto or Lomax) distribution, the Burr type 12 distribution and the F distribution.
They are reasonably tractable, and expressions are given for the pdfs, distribution functions, moments and mode, and random number generation is
also described.

The usefulness of these new distributions might lie in enabling robust analysis, because they resemble their `parent' distributions closely in the main body of the distribution,
whilst being heavy-tailed. 
Thus as the example demonstrated, they can be used for robust analysis in data fitting.
In Bayesian analysis, heavy tailed prior distributions are needed to avoid prior-data conflicts and also to aid efficient computation.
Thus in Markov-chain Monte Carlo (MCMC) we need a proposal distribution that has a similar shape to the target distribution, but a heavy tail.
Such distributions will give good mixing and so aid efficient computation. Giving distributions a heavy tail through the use of the arcsinh transformation
produces distributions that are more similar to the target than those produced by giving the scale parameter a gamma distribution.

\section*{R code}
These functions give the distribution function, pdf, inverse quantile ($x$ value corresponding to
a given value of the distribution function) and random numbers for the arcsinh Weibull distirbution.

\begin{verbatim}
pawb <-
function (x,  beta, nu) 
{
F=1.-exp(-nu*asinh(x**beta/nu))
F
}

dawb <-
function (x,  beta, nu) 
{
xb=x**beta/nu
pdf=beta*x**(beta-1.)*exp(-nu*asinh(xb))/sqrt(1.+xb**2)
pdf
}

qawb <-
function (F,  beta, nu) 
{
x=(nu*sinh(-log(1.-F)/nu))**(1./beta)
x
}

rawb <-
function (n,beta, nu) 
{
if (length(nu)==1) nu=rep(nu, n)
r=runif(n)
x=(nu*sinh(-log(r)/nu))**(1./beta)
x
}
\end{verbatim}
\end{document}